# Le paradigme de la complexité

## Apports pour les approches formelles de l'hypertexte

**Lise Verlaet\*, Sidonie Gallot\* et Audilio Gonzales Aguilar\*\***

*\*LERASS-Céric (EA 827)*
*\*\*IRSIC (EA 4262)*
*Université Paul Valéry*
*Route de Mende - 34 199 Montpellier Cedex 5*
*{lise.verlaet ; sidonie.gallot ; audilio.gonzales}@univ-montp3.fr*

*RÉSUMÉ. Cet article plaide pour un retour aux approches formelles de l'hypertexte et prend appui sur le paradigme de la complexité pour développer l'idée de « site hypermédiateur ». Un site hypermédiateur est un dispositif intermédiaire entre une numérisation de la culture livresque et une « véritable » écriture hypertexte. Si notre réflexion sur les sites hypermédiateurs rejoint les notions d'hypertexte et de bases de données, elle s'en distingue par la relation lecteur-dispositif non plus fondée sur une recherche d'information de type requête mais utilisant l'infovisualisation.*

*ABSTRACT. This article argues for a return to formal approaches of hypertext and builds on the paradigm of complexity to develop the idea of "hypermediator website". A hypermediator website is an intermediate device between a digitalization of book culture and a "real" hypertext writing. If our thinking on the hypermediator website joined the hypertext's notions and the databases, it differs by the relationship reader-device no longer based on information search query but using the visualization of the information.*

*MOTS-CLÉS : approches formelles de l'hypertexte, paradigme de la complexité, systémique, écriture hypertexte, site hypermédiateur, infovisualisation.*

*KEYWORDS: formal approaches of hypertext, paradigm of complexity, systemic, hypertext writing, hypermediator website, visualization of the information.*

**1. Introduction**

Il a fallu plusieurs millénaires pour passer de l'oral à l'écrit, dès son apparition les modes d'échange et de partage d'informations se sont transformés, le savoir et la culture pouvaient enfin laisser une trace, être conservés et transmis aux générations suivantes pour construire, sur ces acquis, de nouvelles connaissances. Car l'écrit est bien plus qu'un simple « mode d'enregistrement de la parole » (Goody, 2007), il a non seulement permis à l'auteur d'user de la rhétorique adéquate pour formuler – et penser – son propos ; mais également aux lecteurs de l'assimiler ou de s'en accommoder (Piaget, 1967) en façonnant ainsi leur réflexion sur le monde qui les entoure. Le travail collaboratif inhérent à la production-appropriation de l'écriture, cette intelligence collective, a indéniablement concouru à l'évolution de la connaissance et, ce faisant, à une (r)évolution de l'espèce humaine.

L'histoire des médias nous montre que dès qu'un nouveau moyen de communication apparaît, une polémique se crée postulant la fin d'un précédent. Tel fut le cas avec l'arrivée d'internet. Or la naissance d'un média n'annonce pas la fin d'un autre, mais entraîne un repositionnement et souvent une évolution des pratiques. Bien que le monde numérique n'en soit qu'à ses balbutiements, internet s'est très vite imposé comme le sixième grand média et a déjà profondément marqué nos rapports à l'information et à la communication. Internet a la particularité de pouvoir réunir à lui seul tous les autres médias. Mais force est de constater que, pour l'heure, ce « roi des médias » semble davantage se faire vampiriser par ses aînés que d'imposer son propre style.

Au sein de cet article, nous revenons sur les différentes approches de l'écriture et de la pensée hypertexte soulignant ainsi le glissement du « concept d'hypertexte » au « lien hypertexte », le passage d'une « technologie de l'intellect » à une « technologie de diffusion », de la « technologie au service de l'homme » à « l'homme au service de la technologie ». Notre positionnement est clair, nous plaidons pour un retour aux sources, cela dit nous sommes également convaincus par l'idée que pour parvenir à une « réelle » écriture hypertexte il soit nécessaire, au préalable, de s'affranchir des normes imposées par le volume. Nous défendons donc l'idée de « sites hypermédiateurs », lesquels constitueraient des dispositifs intermédiaires entre une numérisation de la culture livresque et une « véritable » écriture hypertexte. Si notre réflexion sur les sites hypermédiateurs rejoint les notions d'hypertexte et de bases de données, elle s'en distingue par la relation lecteur-dispositif non plus fondée sur une recherche d'information de type requête mais utilisant l'infovisualisation. En effet, nous soutenons que les représentations des connaissances sont à considérer comme des interfaces prometteuses pour rendre compte d'une réalité complexe et naviguer en son sein.

## 2. « Déviance » des approches formelles de l'hypertexte vers des approches pragmatiques

L'hypertexte est issu d'une longue lignée d'outils conçus pour être des extensions de la mémoire humaine et accroître ainsi nos facultés de raisonnement. À ce titre, il résulte d'un cheminement intellectuel dont les premières traces remontent au 17$^{\text{ème}}$ siècle avec l'invention de la machine à calculer. La problématique supposée de Pascal ou encore de Babbage était d'imaginer un système permettant à l'homme de faire rapidement des calculs complexes sans risque d'erreur. Il s'agit d'externaliser et d'automatiser une partie du raisonnement logico-mathématique d'ores et déjà maîtrisé par l'homme, lui laissant ainsi davantage de temps pour aller plus loin dans sa réflexion et en tester les hypothèses.

### 2.1. *Approches formelles de l'hypertexte*

Les travaux de Vannevar Bush (1945) sont également emprunts de cette idée. Il décrit Memex comme « un supplément agrandi et intime » de la mémoire de son propriétaire, un appareil de stockage d'information à la manière d'une bibliothèque individuelle qui assurerait une consultation facile et rapide. Bush part du principe que le système mnémonique humain repose sur les associations cognitives, le fonctionnement de Memex se devait de calquer le comportement humain afin de tisser des relations entre les documents qui y sont stockés et en permettre la consultation immédiate via un index. Bien que Memex n'ait jamais été réalisé faute de technologie adéquate, il marque un pas important vers un nouveau mode d'accès au savoir, vers des encyclopédies personnelles automatisées.

Si l'on retrouve dans les théories de Bush les prémisses de l'hypertexte, c'est – une quinzaine d'années plus tard – à Ted Nelson que nous devons le terme d'hypertexte (Nelson, 1981, 1993). Avec le concept d'hypertexte, est arguée l'idée d'une écriture résolument non séquentielle laquelle serait coextensive au raisonnement humain. Pour Nelson l'impact culturel de l'imprimerie a façonné notre perception du texte, en associant la présentation d'un texte à une séquentialité. Or, il pointe très justement que la structure de la pensée n'est pas séquentielle par elle-même mais s'assimile à un tissage d'idées (une « structangle »), et par conséquent, la séquentialité d'un texte ne correspond que très rarement à la structure cognitive du lecteur. Seules certaines séquences de texte peuvent intéresser le lecteur dans la construction de son propre système de pensée. L'hypertexte peut, ce faisant, être conçu comme un ensemble de fragments de texte, de granularité variable tant que celle-ci est corrélative à une unité de sens. Cette non-séquentialité doit également permettre de créer différents parcours lesquels seront adaptés aux intentionnalités et au niveau de compréhension des différents lecteurs. Le texte devient dès lors flexible et malléable du fait même de sa non séquentialité. Ainsi, le concept d'hypertexte participe pleinement à l'émergence du sens pour le lecteur car non seulement il lui permet d'accéder à différentes unités de sens mais il révèle également les liens qui les unissent et peut, par là même, répondre aux contraintes d'une réalité et d'une pensée complexe. L'hypertexte apparaît donc comme le concept salvateur du lecteur. Nelson souhaite finaliser sa conception en intégrant l'hypertexte dans un réseau de publication, le projet Xanadu qui finira par aboutir en 1981.

C'est à Douglas Engelbart que l'on doit le premier système fonctionnant sur les principes de l'hypertexte, le NLS (oNLine System). Il partage la même ambition que ces prédécesseurs, celle d'exploiter les capacités des ordinateurs pour augmenter les capacités cognitives et intellectuelles de l'humain. Il voit dans l'interface homme-machine l'occasion pour l'homme d'accroître ses connaissances mais également d'augmenter ses capacités à résoudre des situations problématiques complexes. Pour Engelbart, l'informatique doit apporter aux utilisateurs les informations nécessaires à une meilleure compréhension d'un problème et ce, de façon immédiate. Plus l'utilisateur aura rapidement l'information qu'il recherche pour solutionner son problème, plus il pourra résoudre de problèmes de plus en plus complexes.

Ainsi, nous pouvons observer à travers le concept d'hypertexte cette volonté commune d'imaginer et concevoir des technologies de l'intellect dont l'objectif est d'accompagner l'homme dans sa réflexion et ses raisonnements. Cette vision des pionniers constitue le cœur des approches formelles de l'hypertexte (Rety, 2005). Elle atteint d'ailleurs son paroxysme avec Licklider (1960) lequel pense qu'un jour le cerveau humain et l'ordinateur seront étroitement liés, cette symbiose permettra à l'homme de décupler son intellect. Le bémol – s'il en est – du concept d'hypertexte réside dans le manque d'application ou une marginalisation de ces approches, qui pour certaines relève davantage du fantasme, alors que l'émergence de technologies de plus en plus performantes aurait pourtant dû en favoriser l'expansion.

**2.2.** *Approches pragmatiques de l'hypertexte*

Parallèlement aux travaux sus-cités, d'autres s'affairaient autour des problématiques de transmission de données par ordinateur et de création d'un réseau de communication. C'est, de notre point de vue, cette seconde catégorie de recherches qui a pris l'avantage et c'est sur leur conception que s'est fondé le sens commun du terme hypertexte. Ceci est sûrement dû au fait qu'internet est un système de réseaux interconnectés, descendant d'ARPANET et démocratisé par Tim Berners-Lee via le World Wide Web. Pourtant, les travaux à l'origine du Web se voulaient être un projet de système de gestion de l'information au CERN. Berners-Lee crée dans un premier temps Enquire, système dépeint comme un outil d' « aide intellectuelle » inspiré par les principes de l'hypertexte et utilisé comme support pour se remémorer l'ensemble des interactions entre les chercheurs et leurs projets. Fort de cette expérience, il veut faire évoluer Enquire en un vaste système informatique d'informations en réseau via Internet. Ses principales problématiques ont été de trouver un système, assurant le partage des informations, suffisamment simple d'utilisation de sorte à proposer un outil plus utile que contraignant aux chercheurs, mais surtout compatible avec n'importe quel système d'exploitation utilisé à travers le monde. Le World Wide Web est né avec dans son sillage le langage HTML (HyperText Markup Langage), un langage à balises qui permet de définir la structure logique d'un document avec un ensemble de balises de formatage élémentaires. La simplicité du langage HTML a largement concourue à son appropriation et à son développement.

La facilité d'accès et d'utilisation côté lecteurs, la facilité de création et d'édition côté auteurs accentuée par la multiplication des logiciels d'édition de page HTML et d'environnement WYSIWYG[1], ont suffit pour susciter l'incroyable engouement dont nous sommes aujourd'hui témoins pour cette technologie. Malheureusement cela s'est concrétisé au détriment du concept d'hypertexte. En effet, si l'on compare l'hypertexte tel qu'il est employé à l'heure actuelle avec le concept d'hypertexte soutenu par les approches formelles, nous ne retrouvons – tout au plus – que l'idée de lien entre des documents. Dans la pratique et subséquemment dans la conscience collective, l'hypertexte est une technique qui permet de lier un document à un autre, un objet polymorphe qui permet de naviguer entre des documents. En moins d'une décennie, les liens hypertextes qui jalonnaient le corps des textes se font de moins en moins présents, laissant place à un document écrit de manière séquentielle. L'hypertexte est de plus en plus relégué aux fonctionnalités d'un sommaire. Nous pouvons, dès lors, nous interroger sur les causes de cette « déviance » du concept d'hypertexte vers ce que Rety (2005) qualifie d'approches pragmatiques. Certes la méconnaissance des approches formelles est un facteur, mais il faut également considérer l'emprise de la technologie pour la conception de systèmes d'information. Prenons les logiciels d'édition et en particulier les CMS[2] qui rencontrent un franc succès et étudions succinctement l'exploitation faite de l'hypertexte. Ces technologies proposent différents modèles navigationnels qui participent à l'architecture de l'information, à l'ergonomie de l'interface et assurent le passage d'une page à une autre. Ces modèles navigationnels basés sur l'hypertexte sont donc employés pour faciliter l'orientation au sein de l'espace d'information. L'affordance « lien hypertexte » est, quant à elle, relayée aux outils de mise en style suggérant la fonctionnalité d'un marque page. L'hypertexte s'apparente ainsi davantage à une technologie de gestion de la diffusion d'information – telle une télécommande – qu'à une technologie de gestion de contenu qui la conduirait vers une « technologie de l'intellect ». Rety souligne *« la difficulté qu'il y a à penser les hypertextes indépendamment des technologies qui permettent leur réalisation »* (Rety, 2005, p.2) et, quelque part, l'ascendant que la technologie a sur l'homme réduisant ceux qui l'utilisent à se plier à ses règles.

**2.3. *Méthodes appliquées à l'hypertexte***

Comme le souligne Clément (2007), l'hypertexte est généralement conçu selon deux méthodes : l'hypertextualisation automatique et l'hypertexte inspiré des bases de données. Or ces deux méthodes nous paraissent trop limitées pour penser un véritable dispositif hypertexte en tant que « technologie de l'intellect ».

---

[1] What You See Is What You Get, en français, ce que vous voyez est ce que vous obtenez.
[2] Content Management System, en français, « système de gestion de contenu » qui permettent à chacun de publier des informations sans connaître les langages informatiques sous-jacents.

L'hypertextualisation automatique est par sa nature même insatisfaisante, les machines ne sont ni aptes à créer du sens, ni à le comprendre (Leleu-Merviel, 2003). La recherche « plein texte » ne portant pas sur le sens mais sur l'entité lexicale qui peut être plurivoque (Eco, 1995) n'est pas suffisamment fine pour être efficiente. Qui plus est, cette méthode est génératrice de surcharge cognitive et de désorientation de l'hyperespace d'où les recherches actuelles portées par le W3C[3] pour aller vers un Web sémantique (Berners-Lee et al., 2001), proposition forte intéressante si l'on occulte la complexité technique des couches supérieures de son architecture. Clément soulève également les approches structurelles, linguistiques et statistiques. Les approches structurelles prennent en considération la logique des documents pour en dégager des unités logiques à partir des titres, paragraphes, etc. Dans ces formes les plus évoluées, ces unités logiques reposent sur les formes rhétoriques (arguments, contre-arguments…) repérables par la machine. Les approches linguistiques se centrent sur la proximité sémantique des entités lexicales permettant ainsi de discerner des syntagmes, plus que courant dans notre vocable, et d'en déduire un réseau sémantique à partir des champs lexicaux. Enfin, les approches statistiques consistent à déceler les proximités thématiques. Certaines techniques, telle la clusterisation, sont développées en prenant appui sur une combinaison de cette approche, à l'exemple du méta-moteur de recherche Carrot Search Engine, mais les résultats se révèlent encore peu probants.

Le second genre de méthode pour l'hypertexte s'inspire des bases de données et des recherches menées en gestion de la documentation. Elle fonde son architecture sur trois niveaux. Le premier niveau est la base de données organisée en champs, ces métadonnées recouvrent le plus souvent les éléments du paratexte. Le second niveau est relatif aux concepts exposés dans les documents. Comme le précise Clément « la couche de l'hypertexte conceptuel qui est la plus caractéristique et qui justifie l'appellation de « technologie intellectuelle » car en organisant les données selon un système qui les lie entre elles, elle leur donne sens et produit une information nouvelle qui n'était pas contenue dans la première couche » (2007, p.3). Le troisième niveau s'entend en termes d'interface utilisateur. Nous abondons dans le sens de Clément lorsqu'il souligne les failles de l' « hypertexte base de données », outre un important formatage des données, les informations ne sont accessibles que par un système de requêtes formalisées qui supposent que le lecteur ait une idée relativement précise de ce qu'il vient chercher et la grande majorité de ces systèmes propose des documents dans leur intégralité, donc résolument séquentiels.

Y a-t-il une bonne solution parmi celles ci-dessus ? Prises isolément, sans doute pas, mais conjointement tout semble encore possible tant que l'on cantonne la machine à ce qu'elle sait faire.

---

[3] World Wide Web Consortium, dirigé par Tim Berners-Lee.

**2.4. *Des sites hypermédiateurs***

Avant de discuter plus avant de notre proposition pour concevoir l'hypertexte, il est nécessaire de préciser le cadre de notre étude. Nous nous intéressons à la production scientifique en ligne, et force est de constater qu'en l'occurrence, alors même que nous devrions être un modèle d'exemplarité, les revues et archives en ligne restent claquemurées à une numérisation de la version papier, soumises au carcan du volume et aux normes ancestrales de la culture livresque. Bouleverser ces normes reviendrait à remettre en question notre système de production scientifique, ce qui – sans mauvais esprit – n'est pas près d'arriver. Certes, des exemples prometteurs de documents collaboratifs et exploratoires apparaissent – nous pensons notamment au désormais célèbre texte de Pédauque – mais restent marginaux, et *in fine*, sont assez éloignés de la notion d'hypertexte. Ne soyons pas dupes, les valeurs et normes collectivement partagées mettent du temps à évoluer et à changer. Sans convoquer les concepts inhérents aux théories du changement, nous insistons sur le fait qu'il est nécessaire de procéder par étapes et de montrer à la communauté l'intérêt qu'un changement normatif pourrait occasionner pour l'ensemble des acteurs.

Pour brosser rapidement les contours de nos recherches, nous envisageons l'hypertexte non pas comme un dispositif autonome – du moins dans un premier temps – mais davantage comme un « site médiateur » (Davallon & Jeanneret, 2004). Si ce n'est que dans notre approche, il n'officie nullement comme un portail d'information composé de liens hypertextes qui renvoient vers d'autres sites externes, mais comme un site complémentaire et intrinsèque à une revue ou une archive qui propose un traitement effectif de son corpus pour en dégager un sens nouveau. De par cette distinction, nous préférons le qualificatif de « site hypermédiateur ». Ce traitement inhérent à ce site hypermédiateur ne considère pas les textes intégraux mais les unités de sens portant sur les concepts qui les composent. Les fragments d'informations balisés et traçables sont alors réorganisés pour former de nouveaux documents et sont issus de l'intelligence collective des auteurs présents dans le site-source. Ce nouveau document permet au lecteur de s'informer, de comparer, de confronter les idées, de comprendre les interactions entre les concepts (etc.) et donne ainsi lieu à une réflexion plus poussée mais également plus fine. Si le médiateur qui conçoit l'hypertexte se voit effectuer un travail d'architecte pour assurer la qualité, la complexité et l'adaptativité de ses propositions, il peut prendre appui sur le « cadre  contextualisant » commun pour « extraire » les structures stables. En effet, dans toute société, dans toute culture, il existe une institution sociale et une communauté humaine (Marty, 1990). Il existe effectivement, outre une culture livresque ancrée, une « culture communautaire » plus ou moins partagée de connaissances en fonction des domaines, cette idée de culture communautaire évacue en partie les aspects aléatoires individuels et nous amène à progresser vers une intelligibilité des phénomènes sémiotiques complexes par une contextualisation « des concepts et de leurs liens » dans un cadre commun partagé. Comment alors, établir ce cadre commun stable permettant de donner du sens aux concepts et à leurs interrelations, dans l'hypertexte ?

**3. Pour une approche complexe de l'hypertexte**

« Qu'est-il arrivé ? Il est arrivé que nos moyens d'investigation et d'action laissent loin derrière eux nos moyens de représentation et de compréhension » (Valéry, 1948).

Comme nous l'avons souligné, les approches pragmatiques présentent des limites certaines comparativement aux prémisses du projet initial. Comme l'illustre la citation de Paul Valéry nous avons perdu, dans cette « déviance », la complexité inhérente à nos schèmes de représentation et de compréhension. Cette dissension importante avec les approches formelles de l'hypertexte, nous amène – à l'instar de Clément (2007) – à solliciter le paradigme de la complexité ainsi que la théorie des systèmes complexes pour reconsidérer le concept d'hypertexte.

**3.1.** *Une complexité simplifiée*

Le texte « papier » ou « numérique » peut être défini comme un système simple puisqu'il relève d'une structuration organisée. Le document présente l'organisation que l'auteur a donnée et veut donner à ses idées. Ce « document par intention » (Meyriat, 1978), par ce biais, est un « modèle » d'appréhension prévisible pour un lecteur qui se voit offrir un parcours pré-tracé avec une structure logique. L'usage que nous qualifions de restreint du texte numérique, puisqu'il s'agit généralement d'une transposition du texte papier agrémentée de liens hypertextes pour faciliter la consultation, l'est tout autant. Les potentiels de lecture restent ainsi enfermés dans des normes livresques. Par potentiels, nous entendons des fragments d'information, des unités de sens pouvant être « transclus » (Nelson, 1999), soit « décontextualisés » du document source pour être « recontextualisés » au sein d'un nouveau document avec un sens inédit (Verlaet, 2010). Les usages communs des documents numériques limitent donc les potentiels d'adaptativité complexe de l'hypertexte, qui dans sa fonction initiale prévoit une organisation pluri-dimensionnelle du texte, organisation qui devait libérer le lecteur du volume en lui permettant de construire ses propres parcours. Le lecteur de l'hypertexte « complexe » en est acteur, il active, construit et reconfigure constamment un agencement dont il se fait le co-architecte. Le co-architecte car il élabore son cheminement à partir des liens établis par le « concepteur » du dispositif de médiation, il y a toujours un « maître quelque part : celui qui produit les règles » (Leleu-Merviel, 2003, p.27) Les liens activés par le lecteur lui permettent d'agencer des concepts et des idées qu'il saisit et qui entrent en résonance avec ses connaissances, sa culture, ses enjeux. Evidemment son parcours, dans l'idéal, devrait être libre et adaptatif. Mais force est de remarquer que si **la séquentialité d'un texte ne concorde que rarement à la structure cognitive du lecteur**, il est également probable que le parcours balisé par le concepteur n'y corresponde pas davantage. Il aura cependant l'intérêt de pouvoir ouvrir des champs d'interactions possibles pour le lecteur, et ce sont les processus de reliance qui organisent sa pensée, organisent son parcours, organisent ses connaissances en donnant du sens aux idées.

### 3.2 *Traitement simple pour des données complexes*

Dans l'action de lecture, la triade auteur-texte-lecteur génère un processus conjoint diachronique de construction d'une réalité qui introduit une dimension complexe et imprévisible : l'interprétation. Comprendre un texte c'est l'interpréter, c'est construire (Piaget, 1980). Ces processus de construction se conçoivent dans la « boîte noire » (Watzlawick, 1980) du système de connaissance du lecteur, mais ils s'élaborent sur des « pistes » laissées par l'auteur. Si ces pistes sont figées dans et par le texte « intégral », leur présence permet d'accéder à une structure stable, celle proposée par l'auteur. Cette structure est donc saisissable et prévisible pour un « expert » qui peut extraire les liens « stables » entre les concepts pour les proposer, par la suite, au lecteur. C'est grâce à l'expertise du « baliseur » que le lecteur va pouvoir extraire toute la pertinence des fragments de textes balisés et leurs interconnexions. Ce balisage expert réalisé par un « lecteur professionnel » (Brouilette, 1996) n'est pas une lecture neutre ou naïve d'un document puisqu'il possède un cadre social de référence qui oriente sa manière d'appréhender un texte, lequel assure la qualité scientifique du balisage. Cela dit, le rôle du baliseur est avant tout de faire surgir les unités de sens tout en respectant la pensée de l'auteur. Nous voyons dans les compétences du baliseur – pour relier et catégoriser les unités de sens – un moyen de « mettre en relief et en synergie » les différentes nuances entre les concepts sans les « couper » vraiment des cadres qui en font émerger le sens. À ce titre, cet expert doit donc se livrer à un important travail de contextualisation. Selon cette définition, le baliseur se positionne comme un médiateur entre l'auteur et le lecteur. C'est exclusivement sous cette condition que le processus de simplification de la complexité ne la détruit pas mais en propose une forme intelligible dans laquelle la dualité système-simple (machine) système-complexe (homme) s'aplanit. Homme comme machine traitent des données, l'un d'une manière mécaniste, l'autre d'une manière complexe et l'objectif est de trouver un compromis pour traiter simplement des données complexes et non pas de simplifier des données complexes, à cette fin le rôle du médiateur se situe à un niveau méta pour articuler le tout et l'organiser de manière intelligible et cohérente.

### 4. Organisation et hiérarchisation de la complexité pour l'hypertexte

Le quatuor lecteur-auteur-texte-médiateur constitue un noyau fort pour concevoir un « écosystème textuel » recomposable en vue d'une plus grande adaptativité aux capacités cognitives humaines. Les concepts de reliance (Morin, 1980) et de quasi-décomposabilité (Simon, 1962) pour les « systèmes adaptatifs » (Buckley, 1967) nous semblent essentiels pour penser l'hypertexte puisque ces quatre « acteurs » interagissent à divers niveaux dans l'hypertexte en intégrant tous, dans leurs structures formellement ou informellement, des parties des éléments des systèmes environnants qu'ils partagent et construisent conjointement.

**4.1.** *Systèmes quasi-décomposables et reliance*

Pour Simon (2004), la quasi-décomposabilité des systèmes complexes constitue le fondement de notre compréhension. La complexité, malgré son caractère imprévisible possède une architecture saisissable qui en permet la décomposition (Simon, 1962), si théoriquement tout système complexe « est décomposable », la décomposition doit en préserver la complexité. À cette fin, le choix du découpage doit être relatif aux unités de sens repérables au sein des documents, ces fragments d'information conceptuelle peuvent porter en eux des indices, les traces spécifiant l'articulation possible entre les concepts, leurs interrelations. Les unités d'information explicitant un concept sont alors considérées comme les « éléments primitifs » du système et figurent une structure « simplifiée » (nœuds). Dès lors, tout système complexe présente une structure « formelle » saisissable. Toutefois, il ne s'agit pas de se limiter à mettre au jour une structure, mais pour préserver la dynamique complexe d'un système, il est essentiel s'intéresser aux interconnexions des concepts, aux liens, aux interactions qui relient les éléments structurels, liens qui font qu'ils tiennent ensemble et s'organisent pour être, dans leur globalité, un système-sens. Ce pari repose essentiellement sur l'idée qu'il existe ce que Bachelard (1934) nomme un « idéal de complexité » qui postule l'aptitude de rendre intelligible et de restituer la complexité sans la détruire. Cet idéal, repose de notre point de vue, sur ce que Weaver (1948) qualifie de « complexité organisée », en d'autres termes, l'organisation de la complexité peut être explicitée. Pour rendre compte de la reliance entre les nœuds conceptuels qui forment la structure « formelle », nous nous intéressons aux unités de sens qui caractérisent des relations entre les concepts (interconnexions). Ces relations interconceptuelles catégorisées constituent la structure « informelle » et permettent de révéler le système signifiant. Nous pouvons désormais postuler que pour l'hypertexte, le système complexe recomposé constamment par la triade auteur-texte-lecteur peut être « quasi-décomposée » en s'attachant pour un méta-médiateur – le baliseur – à respecter deux aspects de structuration pour élaborer une organisation architecturale complexe : la structure « formelle » et la structure « informelle ».

Toutefois, pour être au plus près de la « réalité » des phénomènes, les concepts et leurs interrelations doivent faire l'objet d'une pondération pour refléter la dynamique du système. Un traitement statistique basé sur la récurrence des unités de sens exposant les mêmes interrelations entre concepts est donc nécessaire. Cette pondération statistique souligne la « force des interactions » (Simon, 2004) entre des unités de sens interconnectées selon des interactions « fortes, modérées ou faibles » (Simon, *ibid.*), hiérarchisation qui permet de conférer un caractère évolutif, adapté et adaptable aux représentations des connaissances, mais surtout qui reflètent et permettent un processus dynamique en construction permanente dans un cadre pré-structuré par des liens qui font sens. Ainsi, la pondération des nœuds et de leurs interconnexions va déterminer à la fois le système complexe « stable » généré par les interactions fortes mais également les systèmes « instables » corrélatifs aux interactions modérées ou faibles.

### 4.2. *Liens sémiotiques, organisations et recompositions*

Les liens « sémiotiques » apparaissent comme centraux pour penser une approche complexe de l'hypertexte. Le fait que l'hypertexte se présente comme un ensemble « non structuré » *a prioiri* (Clément, 2007) puisque soumis à l'adaptation et au principe d'auto-éco-organisation (Morin, 1990) par le texte lui-même, l'auteur, le médiateur, le lecteur, la finalisation de l'hypertexte est déterminée structurellement au terme d'un parcours dans un ensemble envisagé comme fluctuant. Si, en amont, le médiateur a mis en lumière une organisation, le contexte temporel et les logiques d'organisation dans l'action par le lecteur ne sont pas négligeables. Selon la qualification et le choix des liens, l'action du lecteur sur l'hypertexte structure et construit le système « texte » par l'activation des nœuds, il passe par des phases transitoires et dynamiques via des processus synchroniques et diachroniques (Le Moigne, 1977). Au sein de l'hypertexte « pré-structuré », c'est le lecteur qui active les nœuds et change l'état du système (diachronie) pour parvenir – après avoir construit son parcours – à un état de stabilité (synchronie), chaque activation transforme la structure initiale et change l'état du système. Cette dynamique de l'hypertexte est ce qu'il fait qu'elle se rapproche et devient un support à la pensée du lecteur, une « technologie de l'intelligence ». Cependant « techniquement », cette dynamique repose intégralement sur les liens qui constituent des traces de lectures (laissées par l'auteur), les traces de sens (laissées par le médiateur), ils construisent la structuration de la connaissance, un cheminement de pensée. Ces liens sémiotiques font sens et contextualisent les « textes » consultés, ils sont un « indice », une métacommunication (Palo Alto, 1977) qui qualifie les relations entre des passages de textes. Ces liens qui font sens rendent compte de la complexité, leur hiérarchisation témoigne et organise la complexité.

Le lien proposé devient « signe-passeur » (Le Marec & Jeanneret, 2003) mais plus que cela, il devient « sens », nous parlerons alors de « sémio-trace », et permet au lecteur de structurer son « programme d'activité » (Jeanneret & Davallon, 2004). Le sens du lien permet de conjoindre la technique, l'émergence de sens et la trace de son propre parcours. Le texte qu'il « reconstitue » par l'activation de liens signifiants devient le contexte interprétatif qui le conduira à activer de nouveaux liens qui rétroactivement contextualiseront à leur tour les connaissances et le sens du texte qui l'auront mené ici. Dès lors, le parcours d'activité du lecteur dans l'espace hypertexte fluctuant est hypercomplexe et imprévisible et la difficulté à identifier les intentionnalités pour concevoir et proposer des programmes d'activités, le choix des « sémio-traces » nécessitent une importante expertise du médiateur pour concevoir des parcours adaptatifs et adaptés aux usagers et à leurs « pratiques » cognitives et techniques.

### 5. Représentations collectives, sites hypermédiateurs et visualisation

Si, pour s'approcher de l'idéal des pionniers le retour à une forme complexe de l'hypertexte semble appropriée, la difficulté d'initier un hypertexte complexe est de taille. Car, à constater l'évolution historique des hypertextes, à voir l'usage contemporain, nous pouvons dégager que ce qui a résisté le plus à la mise en œuvre du projet initial c'est précisément la complexité-même des raisonnements humains.

### 5.1. *Concevoir des hypertextes complexes*

Nous allons, pour « penser » comment initier un hypertexte complexe, revenir sur l'idée de structures « stables » de systèmes, c'est à dire des éléments « généraux » récurrents qui constituent une structure relativement stabilisée. Nous avons vu que nous pouvions envisager un système d'organisation « stabilisé » conçu sur la base statistique comme système caractéristique des représentations communes et l'agrémenter de systèmes « instables » prenant en considération les représentations subsidiaires mais existantes. Comme Simon le remarquait à propos des systèmes quasi-décomposables, cette disposition permettrait aux liaisons plus faibles de se transformer sans perturber les liaisons stables des systèmes plus grands, en faisant émerger la hiérarchie sur les « liaisons » de haut niveau (définies par l'intelligence collective, l'auteur, le domaine) et en laissant la liberté aux liaisons « plus faibles » de se transformer en « contexte » dans le système stable et d'éviter ainsi le chaos et la désorganisation.

### 5.2. *S'émanciper du texte pour concevoir l'hypertexte*

Pour penser la représentation des connaissances sur les sites hypermédiateurs, nous revenons brièvement aux modèles de construction des connaissances, de reliance pour dépasser les dualités « éléments-concepts », « liens-sémiotraces » pour les envisager conjointement. À propos de la faculté de reliance, Le Moigne souligne que la « […] faculté de reliance […] nous épargne les navigations cognitives qui vont du réductionnisme qui sépare, au holisme qui fusionne ; la reliance révèle et organise des « patterns » d'interactions possibles par lesquelles les « complexes » nous deviennent intelligibles […]» (Le Moigne, 2008, p 181). La reliance s'opère par les liens qui interfacent les idées par la conception opérationnelle de patterns qui rendent intelligibles les « interactions complexes », le pattern relevant d'une faculté naturelle de l'esprit.

Pour Hanson (1958), les patterns sont des modèles de découvertes que nous construisons et qui nous permettent d'organiser nos connaissances pour « comprendre », il le définit comme « la chose qui fait que la chose soit la chose ». Le pattern est donc un outil de découverte qui contribue à construire des connaissances par reliance, ils sont des modèles organisateurs (Simon, 1979 ; 1981) qui rendent compte des comportements et des évolutions des phénomènes. Bateson (1979) également parle des « patterns which connects », c'est-à-dire d'une reliance entre « le corps », « l'esprit » et « le contexte » comme à même de rendre compte du tout et des parties à travers des schèmes récurrents complexes mais identifiables parce que récurrents, incluant l'individu et son esprit dans le processus. Encore et surtout le concept de pattern revêt, notamment dans les arts graphiques, l'idée figure d'un « dessin » récurrent, reconnaissable et reproductible.

Pour Simon, ces patterns relèvent d'un processus intrapsychique complexe qui se construit sur des « traces », c'est à dire sur une « réserve mémorisée » qui permet de comprendre de nouvelles situations, de nouveaux phénomènes en servant de « cadre conceptualisant ». Cette idée de trace sémiotique nous semble intéressante si nous l'envisageons comme des indications sur les « parcours » de lecture pour l'hypertexte, c'est pour cela que nous parlons de sémio-traces. Si ces traces sont « personnelles », comme nous l'avons évoqué, l'idée de culture commune, l'appel à l'intelligence collective, à la représentation visuelle semblent à même de nous guider vers une émancipation du carcan du volume. En effet, penser à mettre à profit l'intelligence collective pour collecter les « sémio-traces » des individus, sémio-traces récurrentes « communes » faisant sens pour le plus grand nombre permettrait de penser des liens plus complexes, plus fins, mais surtout plus signifiants et stables. Il est indéniable que dans une telle perspective la multiplication des points de vue pour palier l'idée d'une vision surplombante générale permet de représenter des parcours adaptés à des « communautés » de lecteurs. L'essence de la culture qui va permettre de donner du sens aux liens réside dans « le collectif » dès lors, la mise à profit de l'intelligence collective comme outil pour organiser, hiérarchiser, penser l'hypertexte dans une dimension complexe nous semble tout à fait appropriée.

Pour s'émanciper des normes de l'ouvrage et s'attacher à la mise en œuvre de l'hypertexte telle qu'elle a été fantasmée par les pionniers nécessiterait certainement de briser les schèmes normatifs sclérosants et simplifiants, qui concourent à l'enfermer dans les normes du volume, en ce sens l'idée de représentation mentale, de visualisation, de lisibilité ne constituerait-elle pas une clé en vue d'une adaptation à un plus grand nombre de lecteurs et à leurs représentations individuelles et collectives ? Si, comme l'aurait affirmé Confuscius « une image vaut mille mots », il convient de se demander comment la représentation « visuelle » peut elle devenir un moyen de « libérer » l'hypertexte du volume et de la technique par des représentations des systèmes, des architectures complexes, des concepts, des liens « dynamiques » porteurs de sens si ce n'est par les « représentations » visuelles qui seraient des propositions de parcours de lecture et agiraient alors comme des patterns dont l'assemblage construirait un système cadre contextualisant complexe ?

## 6. Conclusion

Nous avons souligné dans cet article la déviance opérée entre les approches formelles de l'hypertexte initiées par les pionniers. Alors qu'ils les envisageaient comme des « technologies de l'intellect » permettant à l'homme de transcender ses capacités cognitives elles ont glissé vers des approches pragmatiques cloisonnées par la technique et subordonnées par le besoin de communiquer et de diffuser de l'information. Peinant à trouver leurs marques et toujours sous la coupe de la culture livresque, les méthodes de l'hypertextualisation automatique ne peuvent être satisfaisantes, les machines étant dans l'incapacité de créer et de comprendre la complexité humaine. L'émergence du sens ne peut être opéré que par des êtres pourvus d'intelligence, ce qui est le cas pour l'hypertexte inspiré des bases de données mais dont les méthodes trouvent leurs limites dans le système de recherche d'information par les requêtes qu'elles sous-tendent. Nous proposons donc une méthode hybride, prenant ancrage sur les approches complexes pour repenser le concept d'hypertexte et développons l'idée de site hypermédiateur. Les sites hypermédiateurs constituent un intermédiaire et une première étape vers de nouvelles formes de technologies de l'intelligence, vers une intelligence collective d'un monde de connaissance. Les systèmes complexes émanant du balisage du corpus du site-source peuvent être représentés via les techniques de l'infovisualisation et être utilisés pour parcourir l'espace d'information. Goody (2007) décrit l'écriture comme un « mode d'enregistrement de la parole », l'infovisualisation peut – selon nous – être perçue comme un mode d'enregistrement de la cognition.

## 7. Bibliographie

Bachelard G,. *Le nouvel esprit scientifique.* Paris, PUF, 1934, vol. 1.

Bateson G., *Mind and nature : a necessary unit*, Batam books, 1979.

Berners-Lee T., Hendler J., Lassile O., « The semantic Web », *Scientific America*, mai 2001.

Brouilette C., «Vers une définition de la lecture professionnelle », *Cursus*, vol. 1, n°2, 1996. En ligne : http://www.ebsi.umontreal.ca/cursus/vol1no2/, consulté le 13/08/2007.

Buckley W., *Sociology and modern system theory,* New Jersey, Prentice Hall Inc, 1967.

Bush V., "As we may think", *The atlantic monthly*, 1945.

Clément J., « L'hypertexte, une technologie intellectuelle à l'ère de la complexité », *in* Claire Brossaud et Reber Bernard (dir.), *Humanités numériques 1. Nouvelles technologies cognitives et épistémologie*, Hermès Lavoisier, 2007. En ligne : http://www.hypertexte.org/blog/wp-content/uploads/2009/01/techn_intellcomplexitejclement.pdf

Davallon J., Jeanneret Y., « La fausse évidence du lien hypertexte », *Communication et langages*, vol. 140, n° 1,2004, p.43-54.

Eco U., *Interprétation et surinterprétation*. PUF, 1995.

Goody J., « L'oralité et écriture », *Communication et langages*, vol.154, n°1, 2007, p.3-10.

Hanson N., *Patterns of discovery*, Cambridge University Press, 1958.

Le Moigne J.L., « Edgar Morin, le génie de la reliance », *Synergies Monde,* n°4, 2008, p.177-184.

Le Moigne J.L., *in* les Introuvables en langue française de H.A. Simon. En ligne http://www.mcxapc.org/nc/fr/documents/les-documents-complexite-en-oeuvre/les-introuvables-de-ha-simon.html, consulté le 27/06/2011.

Leleu-Merviel S., « Le désarroi des « Maîtres du sens » à l'ère numérique », *H2PTM'03 Hypertextes et Hypermédias, Créer du sens à l'ère numérique.* Paris : Hermès, 2003, p.17- 34.

Licklider J.C.R., « Man-computer symbiosis », *Human Factors in Electronics, IRE Transactions on*, 1960, nº 1, p.4-11. En ligne http://ieeexplore.ieee.org/xpls/abs_all.jsp?arnumber=4503259

Marty R., *L'algèbre des signes*. Amsterdam: John Benjamins, 1990.

Meyriat J., «Document, documentation, documentologie », Schéma et schématisation, n° 14, 1981, p.51-63.

Morin E., *Introduction à la pensée complexe*, Paris : Seuil, 1990.

Nelson T.H., Dream machines. *South Bend,* In *the distributors*, 1974.

Nelson T.H., *Literary Machines,* Mindful Press, 1981.

Nelson T.H., Xanalogical structure, needed now more than ever: parallel documents, deep links to content, deep versioning, and deep re-use. ACM Computing Surveys (CSUR), vol. 31, n° 4, 1999, p.33.

Piaget J., *Psychologie de l'intelligence*. Collection Agora, Editions Armand Colin, 1967.

Piaget J., *La construction du réel chez l'enfant*, Neuchâtel : Delachaux-Niestlé, 1980.

Réty J-H., « Ecriture d'hypertextes littéraires: approche formelle et approche pragmatique », *Actes de H2PTM 2005 Paris, créer, jouer, échanger: experiences de réseaux*, 2005. En ligne http://webperso.iut.univ-paris8.fr/~rety/publis/2005h2ptm.pdf

Simon H.A., *L'unité des arts et des sciences : la psychologie de la pensée et de la découverte*, présentée à l'Académie des Arts et des Sciences des USA, mai 1981.

Simon H.A., *Models of thought, vol I & II,* Yale, Yale university press, 1979.

Simon H.A., *The Sciences of the Artificial*. Cambridge: MIT Press, 1996.

Simon H.A., « Sur la complexité des systèmes complexes », Université Carnegie-Mellon, Pittsburgh, 1971.

Souchier E. et al., *Lire, écrire, récrire : objets, signes et pratiques des médias informatisés.*Paris: éditions de la BPI, 2003.

Valery P., (1988). in "Vues", Ed La table Ronde, 1948, p.41.

Verlaet L., Stratégie de balisage générique pour des bases de données contrôlées : le modèle par l'ASCC. In Amos David, dir. *L'Intelligence Économique dans la résolution de problèmes décisionnels*, collection Hermès Sciences Publications, Edition Lavoisier, Chapitre 8, 2010, p.187-212.

Watzlawick P, *Le langage du changement – Paradoxes et psychothérapie*, Paris : Seuil, 1980.

Weaver W., Science and complexity, *American scientist*, n°36*,* 1948, p.536-544.